\documentclass[prd,twocolumn,showpacs,preprintnumbers,amsmath,amssymb]{revtex4-1}

\usepackage{graphicx}
\begin{document}

\def\gf{\mathfrak{B} }
\def\lf{\mathfrak{L} }

\def\tc{\lambda^t}

\title{Complex RG flows for 2D nonlinear $O(N)$ sigma models}
\author{Y. Meurice}
\author{Haiyuan Zou}
\affiliation{Department of Physics and Astronomy\\ The University of Iowa\\
Iowa City, Iowa 52242, USA }
\date{\today}
\begin{abstract}
Motivated by recent attempts to find nontrivial infrared fixed points in 4-dimensional lattice gauge 
theories, we discuss the extension of the  renormalization group (RG) transformations to complex coupling spaces for  $O(N)$ models on $L\times
 L$ lattices, in the large-$N$ limit. 
We explain the Riemann sheet structure and singular points of the finite $L$ mappings between the mass gap and the 't Hooft coupling. 
We argue that the Fisher's zeros appear on ``strings" ending approximately near these singular points. We show that for the spherical model at finite $N$ and $L$, the density of states is stripwise polynomial in the complex energy plane. 
We compare finite volume complex flows obtained from the rescaling of the ultraviolet cutoff in the gap equation and from the two-lattice matching. 
In both cases, the flows are channelled through the singular points and end at the strong coupling fixed points, however strong scheme dependence appear 
when the Compton wavelength of the mass gap is larger than the linear size of the system. We argue that the Fisher's zeros control the global properties of the complex flows.
We briefly discuss the implications for perturbation theory, proofs of confinement and searches for nontrivial infrared fixed points in models beyond the standard model.

\end{abstract}
\maketitle

\section{Introduction}
Recently, there has been a renewed interest 
\cite{Sannino:2004qp,Sannino:2009za,shamir08, appelquist09, hasenfratz09,fodor09, Deuzeman:2009mh,Myers:2009df,DelDebbio:2010ze}
in the possibility \cite{banks81} of finding nontrivial infrared (IR) fixed points in asymptotically 
free gauge theories with enough matter fields. A particularly interesting situation is when in addition of the nontrivial IR fixed point one ultraviolet (UV) fixed points also appears at larger coupling.  
It has been argued 
\cite{Kaplan:2009kr,moroz09} that in this type situation, a parameter can sometimes be varied in such  a way that these two fixed points coalesce and then disappear in the complex plane. 

This observation has motivated us 
\cite{Denbleyker:2010sv} to study complex extensions of renormalization group (RG) flows in the complex coupling plane. The main 
result is that the Fisher's zeros - the zeros of the partition function in the complex coupling plane - act as a ``gate" for the RG flows ending 
at the strongly coupled fixed point. This can be seen as a complex extension of the general picture proposed by Tomboulis \cite{Tomboulis:2009zz} to prove confinement: the 
gate stays open as the volume increases and  RG flows starting in a complex neighborhood  the 
UV fixed point (where we have asymptotic freedom) may reach the IR fixed point where confinement and the 
existence of a mass gap are clearly present. 

More generally, constructing RG flows in complex spaces could improve our understanding of the convergence of expansions (such as weak coupling and strong coupling expansions) that are used in the neighborhood of fixed points. Even though, complexification is often used for hydrodynamical  flows, 
we are only aware of two previous studies of complex RG flows:  
one for exactly solvable lattice models \cite{Damgaard:1993df} and one discussing the possibility of chaotic behavior in the decimation of one-dimensional Ising models with complex coupling \cite{PhysRevE.52.4512}. 

In the following we discuss two complex extensions of RG flows for  $O(N)$ models on $L\times L$ lattices, in the large-$N$ limit. 
The  models are introduced  in Sec. \ref{sec:model}. 
 We provide a closed form 
expression for the partition function in the approximation where the 
non-zero modes of the Lagrange multiplier are neglected. This is justified in the large-$N$ limit where we have equivalence with the spherical model. 
 In Sec. \ref{sec:gap}, we study the map between the mass gap $M^2$ and the 't Hooft 
coupling $\tc =1/b$. We show that the 
map requires a Riemann surface with $q+1$ sheets  and $2q$ cuts in the 
$\tc$ plane, where $q$ is an integer of order $L^2$. By connecting the 
sheets in a specific way, we construct one circle at infinity in the $\tc$ plane (or around 0 in the $b$ plane) that maps 
into the circle at infinity in the $M^2$ plane and $q$ others that maps in small regions near 
real interval $[-8,0]$. 

In Sec. \ref{sec:zeros},
we  use the closed form of the partition function of Sec.  \ref{sec:model}
to calculate the Fisher's zero at finite $L$ and $N$.
We show empirically that these zeros appear on ``strings'' coming from 
infinity in the $b$ plane and ending near the singular 
points of the map discussed in Sec. \ref{sec:gap}.  This is consistent with the infinite volume picture provided in Ref. \cite{Meurice:2009bq}.
The density of zeros on these strings 
scales like $L^2$ and $N$. The results of this section can be compared to what is found for other models, for instance in Refs. \cite{alves90b,janke01,janke04,Janke:2002ps} where one and two-dimensional structures have been observed. 

In Sec. \ref{sec:density},
we show that the density of states is 
piecewise polynomial on $q$ horizontal strips in the energy 
plane. We discuss the conjecture that connects the Fisher's zeros with 
the zeros of the second derivative of the logarithm of the density of 
states. In Sec. \ref{sec:flow}, 
we extend two RG methods to the complex $b$ plane. The first one is based on a simple rescaling of the cutoff in the gap equation. The second one is 
a procedure called the two-lattice matching  \cite{PhysRevB.27.1736,Hasenfratz:1984hx} . 
All the numerical calculations are done with $L$ and $N$ even. 

Before embarking into the technical discussion, it is important to keep in mind our various motivations.  From  a practical point of view, it is easier to calculate Fisher's zeros than to construct RG flows and establishing a clear connection should provide more robust ways to decide 
about the existence of nontrivial IR fixed points.
The more general question of understanding the analytic properties of the map between the coupling and the mass gap in the complex plane is also important and includes the correction to asymptotic scaling. This question can be studied explicitly for the models considered here and the results illustrate the intricate pattern that can be produced 
by combining lattice artifacts and finite size effects. Possible implications for lattice gauge theory will be discussed in the conclusions. 
\section{The model}
\label{sec:model}

The partition function for the  $O(N)$ nonlinear sigma model on a square, or more generally hypercubic lattice with volume $V=L^D$ reads:
\begin{equation}
\label{eq:pf}
Z=C\int \prod _{\mathbf x} d^N\phi_{\mathbf x}\delta(\vec{\phi}_{\mathbf x}.\vec{\phi}_{\mathbf x}-1) {\rm e}^{-(1/g_0^2)\sum_{{\mathbf x},{\mathbf e}}(1-\vec{\phi}_{\mathbf x}.{\vec{\phi} }_{\mathbf x+e}) }\ ,
\end{equation}
We denote the inverse  't Hooft coupling as $b\equiv1/\tc\equiv 1/{g_0^2N} $. The constraint $\vec{\phi}_{\mathbf x}.\vec{\phi}_{\mathbf x}=1$ can be implemented by introducing a Lagrange multiplier $M^2_{\bf x}$ at every lattice site. After this is done, the action becomes quadratic in $\vec{\phi}$ and the Gaussian integration can be performed. It can be shown  that in the large-$N$ limit only the zero mode of the Lagrange multiplier, denoted $M^2$ hereafter, survives \cite{Polyakov:1987ez}. With this simplification, the partition function becomes:
\begin{equation}
\label{eq:bkpf}
Z(b) =\frac{\Gamma(\frac{NV}{2})}{2\pi i ({\frac{bNV}{2}})^{\frac{NV}{2}-1}} \oint_C dM^2 e^{\frac{VN}{2}[ bM^2-\lf\left(M^2\right)]}
\end{equation}
with
\begin{equation}
\lf\left(M^2\right)=\frac{1}{V}\sum_{{\bf k}}\ln[2\sum _{i=1}^D \left(1-\cos \left(k_i\right)\right)+M^2]\ .
\end{equation}
The contour of integration, denoted $C$, encircles the real interval $[-8,0]$, also called ``the cut". When $Reb>0$, $C$ can be deformed into a vertical line with an arbitrary positive real part and a semi-circle at infinity going counterclockwise from $\pi/2$ to $3\pi/2$ which gives no contribution in the limit of infinite radius. Similarly,  when $Reb<0$, $C$ can be deformed into a vertical line with a negative real part smaller than -8 and a semi-circle at infinity going clockwise from $\pi/2$ to $-\pi/2$. The prefactor has been adjusted in order to have $Z(0)=1$ and the finite volume momenta take the values ${\bf k}=\frac{2\pi}{L} {\bf n}$  with ${\bf n}$ a vector of integers modulo $L$. 

For $N$ even, the exponential of $-V(N/2)\lf\left(M^2\right)$ is a product of poles located at the real negative values 
\begin{equation}
\label{eq:listpoles}
 M^2_j=-2\sum _{i=1}^D \left(1-\cos \left(k_i\right)\right)\ .
 \end{equation}
The integer $j$ indexes the various values taken while ${\bf k}$ runs over its $V$ possible values ${\bf k}=\frac{2\pi}{L} {\bf n}$. The number of distinct poles will be discussed in section \ref{sec:gap} where it is denoted $q+1$. The number of times a given value of $M^2_j$ occurs will be denoted $n_j$ and we have $\sum_j n_j=V$.  

By calculating the residues, we get a general expression of the form
\noindent
\begin{equation}
Z(b)=\sum_{i,j}{a_{ij}(\frac{1}{b})^i e^{\frac{V N b M^2_j}{2}}}\ ,
\label{eq:bk}
\end{equation}
where $a_{ij}$ are coefficients depending on the order $i$ and the pole $M^2_j$.  For a given $j$, the pole is of order $n_jN/2$ and the index $i$ 
runs between $VN/2-1$ and $(V-n_j)N/2$. 
An explicit expression for  $D=2$, $L=4$, $N=2$ is given in Eq. (A.1) in the Appendix. Despite the apparent singularities at $b=0$, $Z(b)$ is an entire analytical function and has a regular expansion at $b=0$. For instance for  $D=2$, $L=4$, $N=2$
\begin{equation}
Z(b)=1-64 b+\frac{35328 b^2}{17}-\frac{2326528 b^3}{51}+O\left(b^4\right)
\end{equation}

If in addition $L$ is even,  then for every $M^2_j$, there is an associated $M_{j'}^2=-8-M_j^2$ obtained by changing all the $k_i$ into $\pi -k_i$ and one can see that $a_{ij'}=(-1)^ia_{ij}$. This guarantees that 
\begin{equation}
\label{eq:dual}
Z(-b)={\rm e}^{b4VN}Z(b)
\end{equation}
 as explained in \cite{Meurice:2009bq}.
 
It should be noted that the number of independent $a_{ij}$ grows like $L^2 \times N$. This proliferation of terms makes calculations 
performed in the next sections slow when $L$ or $N$ becomes too large. For illustrative purpose, we will often use $L=4$ and $N=2$. 
This allows us to give explicit  formulas of decent size as in the Appendix. However, it should be kept in mind that 
Eq. (2) is only a good approximation of the original partition function  (Eq. (1)) for large $N$. 

\section{The Gap equation, Singular Points, and Cuts}
\label{sec:gap}
\begin{figure}[h]
 \begin{center}
	\includegraphics[width=3.5in]{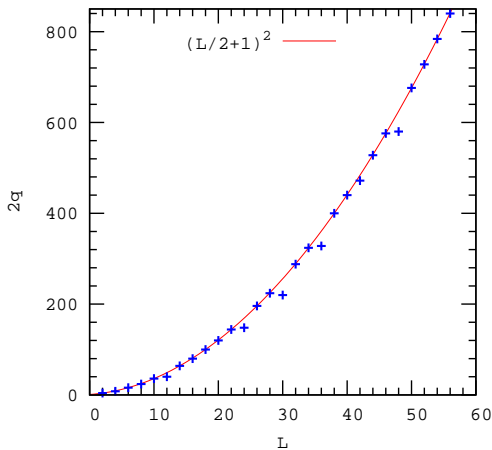}
\end{center}
		\caption{\label{fig:number} Relation between $L$ and $2q$ for $D=2$. 
}
	\end{figure}

In the large-$N$ limit, it is possible to calculate the partition function in the 
saddle point approximation. 
Varying $M^2$, we obtain the gap equation: 
\begin{equation}
\label{eq:gap}
b=d\lf(M^2)/dM^2\equiv\gf(M^2)\ ,
\end{equation}
with \begin{eqnarray}
\gf(M^2)&\equiv&(1/V)
\sum_{\bf k}\frac{1}{2(\sum_{i=1}^D(1-{\rm cos}(k_i))+M^2}\\
&=&(1/V)\sum_{j=1}^{q+1}\frac{n_j}{M^2-M^2_j}
\end{eqnarray}
where $n_j$ is the number of times the pole $M_j$  appears in the sum over the ${\bf k}$ and $q+1$ the number of distinct poles. 
The  explicit form of $\gf(M^2)$  in the case $D=2$ and $L=4$ is given in Eq. (A.2) in the appendix. From now on  $D=2$ is assumed in all the examples. 
The precise value of $q$ depends on accidental degeneracies but  generally increases like $L^2$ . 
From Fig. \ref{fig:number} and Table \ref{tab:n}, we see that most of the values of $2q$ follow the relation $(L/2+1)^2$ or $(L/2+1)^2-1$ with exceptions every three or six data points.
In general, after reducing to a common denominator, we obtain a rational form :
\begin{equation}
\gf(M^2)=Q(M^2)/P(M^2)\ ,
\end{equation}
where $Q$ and $P$ are polynomials of degrees $q$ and $q+1$ respectively.  

\begin{figure}
\includegraphics[width=2.5in,angle=270]{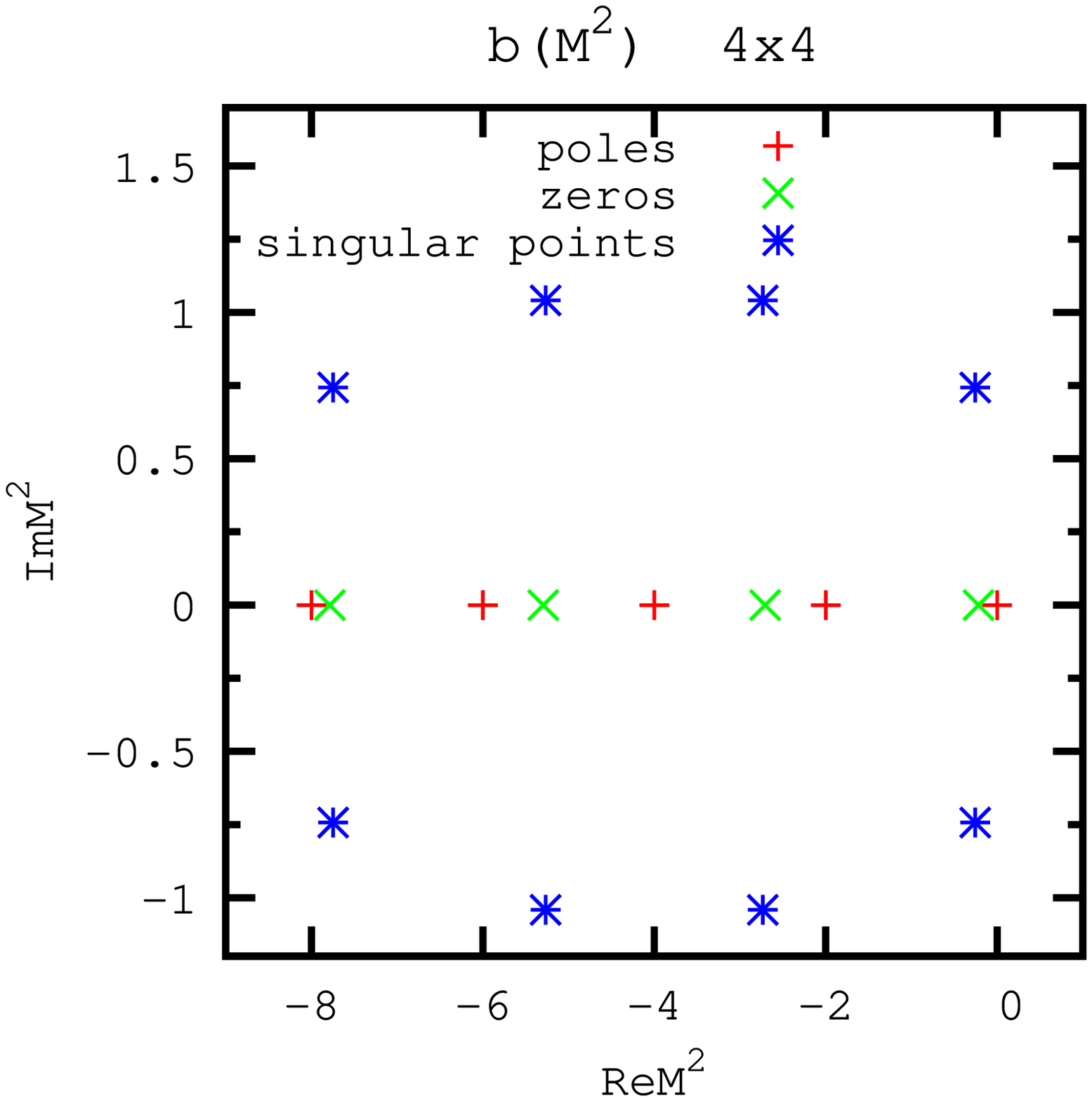}
\includegraphics[width=2.5in,angle=270]{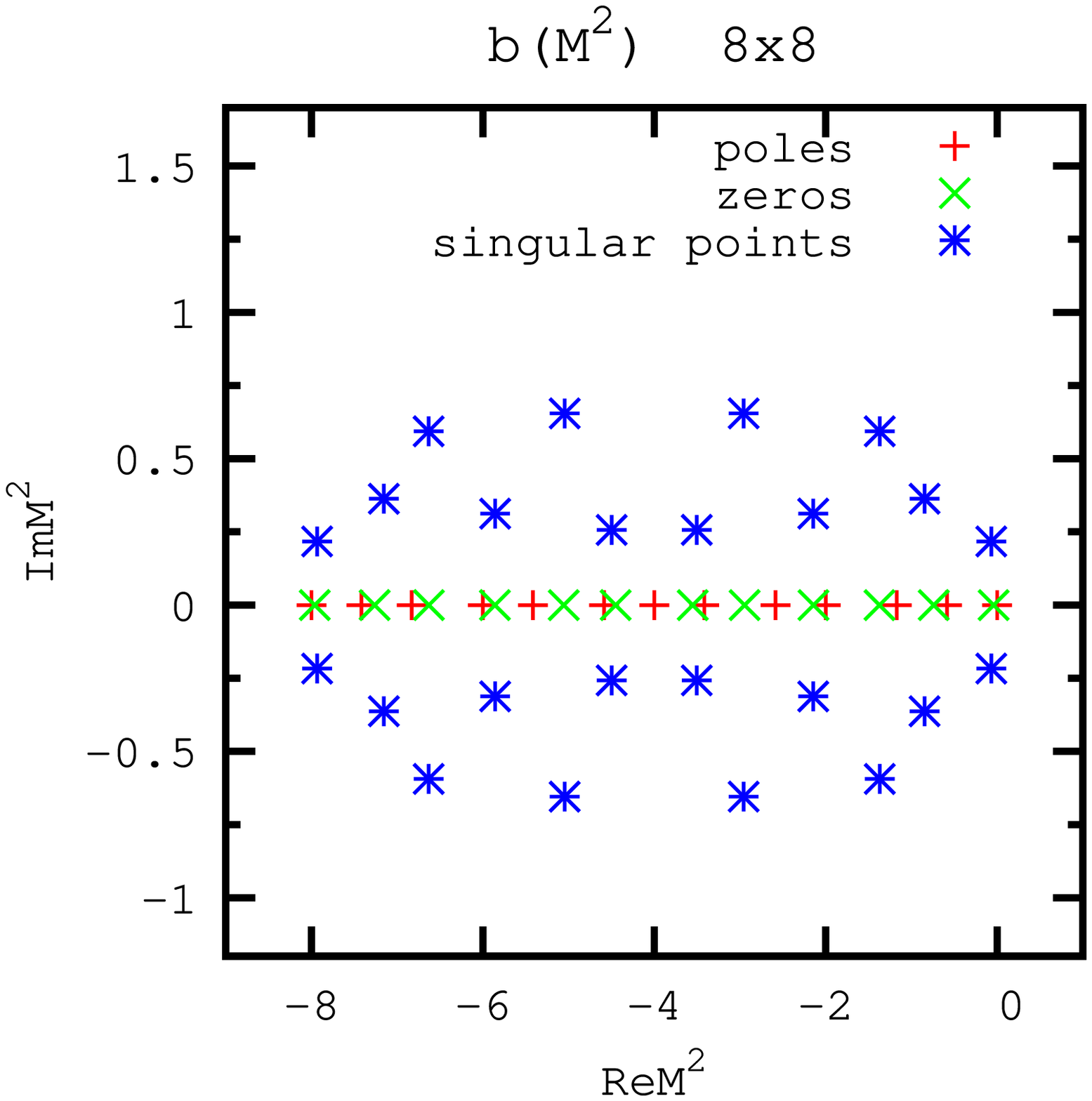}
\caption{\label{fig:fvmap} Zeros, poles and singular points  of $b(M^2)$ in the $M^2$ plane for 4$\times$4 and  8$\times$8 lattices.}
\end{figure}

We now discuss the poles, zeros and singular points of the mapping between $b$ and $M^2$ given by the gap equation (\ref{eq:gap}). 
From Eq. (\ref{eq:listpoles}), the $q+1$ poles of $\gf$ are real and between -8 and 0. -8 and 0 are always poles and on the real interval between them (that we call ``the cut'' hereafter),  
$\gf(M^2)$ is zero once between each pair of successive poles. In addition of these $q$ zeros, 
$\gf(M^2)$ is also zero  when $M^2$ becomes infinite. This also makes $q+1$ zeros. In general, $b=\gf(M^2)$ takes all the complex values $q+1$ times  when $M^2$ is varied over the whole complex plane. Thus, the inverse map between the mass gap $M^2$ and  $b$ requires a Riemann surface with $q+1$ sheets in the $b$ plane.   
To decide where to put the cuts and how to join different sheets, we need to study the singular points where $\partial b/\partial M^2 =0$.  This occurs when $P'Q-PQ'$, a polynomial of degree $2q$, vanishes. The $2q$ roots of $P'Q-PQ'$ appear in complex conjugate pairs in $M^2$ plane. This is illustrated  in Fig. \ref{fig:fvmap} for $L$ = 4 and 8. We notice that as $L$ increases the region where the singular points appear shrinks along the cut. A log-log plot of the largest imaginary part of the singular points 
versus $L$ is rather irregular but suggests that the height of the region where the singular points appear is of the order $1/L$. In the infinite volume limit, the singular points become dense and cover the $[-8,0]$ cut.

\begin{figure}[h]
 \begin{center}
\includegraphics[width=3.5in]{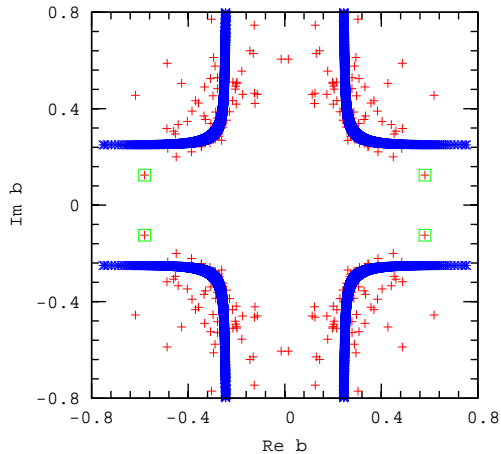}
\end{center}
\caption{
\label{fig:l32}The blending
small crosses (x, blue online) are the $b$ images of two lines of
points located very close above and below the $[-8, 0]$ cut in infinite volume;
the crosses (+) are the images of the singular points for $L=32$. The
images of the 4 closest singular points appear as boxes.
}
\end{figure}

The image of a singular point $M^2_{sing.}$ in the $b$ plane is $\gf(M^2_{sing.})$. At infinite volume,  $\gf(M^2)$ becomes an integral with four logarithmic singularities \cite{Meurice:2009bq}. The image of two lines of points located very close above and below the $[-8, 0]$ cut, span four curves forming a cross shaped figure that can be seen in Fig. \ref{fig:l32}. For comparison, the 288 singular points for a 32 $\times$ 32 lattice are also displayed. We find that the real part of the closest singular points (CSP) move to infinity while the imaginary part stay at $\frac{1}{8}$ as the volume increases. Near a singular point $M^2_{sing.}$, we have $(\gf(M^2_{sing.}+\delta z) -\gf(M^2_{sing.}))\propto (\delta z)^2 $ and we need two sheets to invert the function in the neighborhood of $\gf(M^2_{sing.})$. 

\begin{figure}[h]
	\includegraphics[width=3.2in]{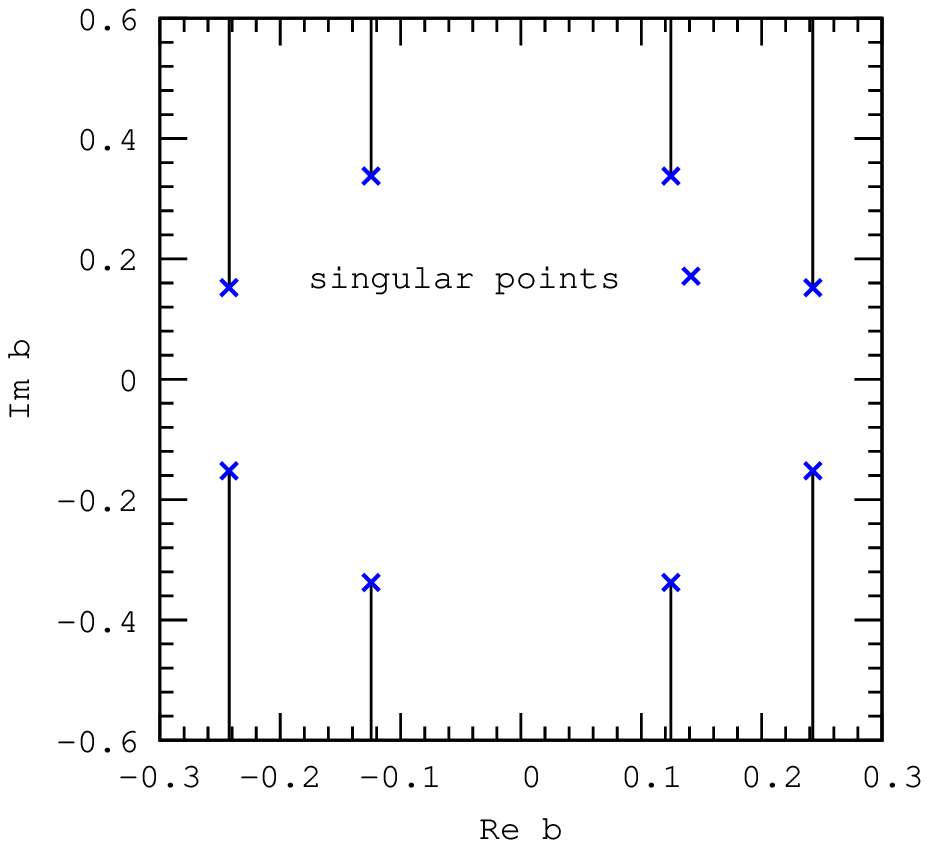}
		\includegraphics[width=3.2in]{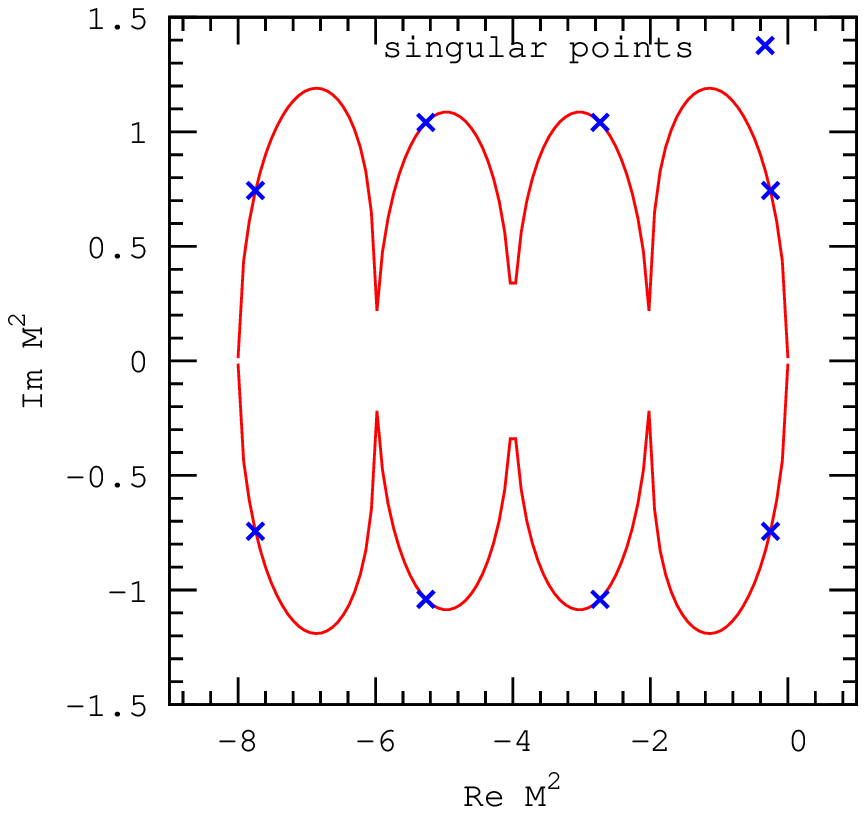}
		\caption{
\label{fig:cut}Singular points and cuts in the $b$ plane for $L=4$ (top) and their inverse images in the $M^2$ plane. 
}
\end{figure}

In order to construct the $q+1$ sheets, we start with the region of the $M^2$ plane where $|M^2+4|>>8$. In this region we have $b\simeq 1/M^2$. We call this sheet the ``main'' sheet because it contains the usual strong coupling region where $b$ is small, real and positive corresponding to a $m_{gap}^2$ large, real and positive. As we now consider smaller values of $|M^2+4|$, and correspondingly larger values of $b$ on the main sheet, we start running into singular points and need to decide on the location of the cuts. A simple choice is to take the cuts on vertical lines in the $b$ plane going from the images of singular points with positive imaginary part to 
increasing values of the imaginary part and from the images of singular points with negative imaginary part to 
decreasing values of the imaginary part. The cuts are shown for $L=4$ in Fig. \ref{fig:cut}. We can now construct the inverse image of the two branches of a cut on the main sheet. They end up on two real negative values of $M^2$ where $b$ becomes infinite. The complex conjugate of the inverse image of these two branches corresponds to the complex conjugated cut in the $b$ plane.
Joining the two together, we obtain an oval shaped region in the $M^2$ plane located symmetrically across the cut. If we vary $M^2$ inside each of the oval shapes, $b$ runs over the whole complex plane forming the other $q$ sheets. 
This construction is illustrated for $L=4$ where $q=4$ in Fig. \ref{fig:cut}. 
If a curve in the $M^2$ plane  enters an oval shaped region say on the left of the critical point and exits on the right of this critical point, then its image in the $b$ plane will wrap around the image of the singular point. In Section VI, we will show that the cuts in the $b$ plane and the boundaries of the oval shaped regions are important to understand  the RG flows. 

\section{Fisher's Zeros}
\label{sec:zeros}
\begin{figure}[h]
 \begin{center}
	\includegraphics[width=3.9in]{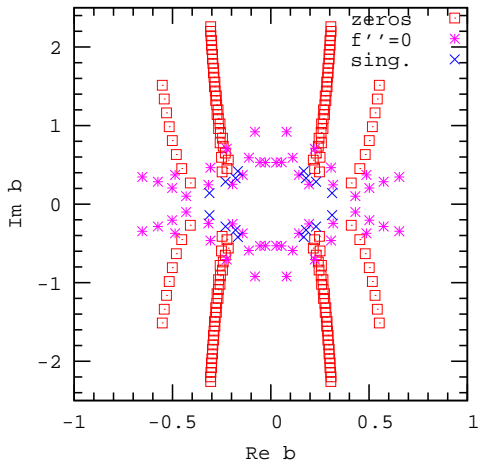}
\end{center}
		\caption{
				\label{fig:zero6} Zeros of partition function for $L=6$, $N=2$ (boxes), and images of the singular points of $b(M^2)$ (crosses). The images of the solutions $f''=0$ discussed in section \ref{sec:density} are given with the third symbol
}
\end{figure}

In this section, we discuss the Fisher's zeros of the partition function at finite $N$ and $L$ and the way their density scales with these quantities. 
Later we will show that these zeros play an important role in controlling the RG flows. 
The coefficients $a_{ij}$ in the large-$N$ expression Eq. (\ref{eq:bk}) can be calculated {\it exactly} at finite $N$ and $L$ using the residues theorem. We can then 
search for the zeros of the partition function by using Newton's method and check that  the number of zeros found inside a given region of the $b$ plane encircled by a closed curve $C$
is consistent with
\begin{equation} 
\oint_C  db (dZ/db)/Z = 
i2\pi \sum_q q n_q(C)\ ,
\end{equation}
where $ n_q(C)$ is the number of zeros of order $q$ inside $C$. The results are shown in Fig. \ref{fig:zero6} for $L=6$. We see that the zeros form linear structures (``strings") ending at locations close to the ($N$-independent) singular points. Similar pictures are found for other not too large values of $L$ and $N$ where similar calculations are feasible. In all the examples considered, we also found that the zeros closest to the real axis always have an imaginary part larger than 1/8 in absolute value, in other word they never get closer to the real axis than the CSP. 

\begin{figure}
	\includegraphics[width=3.5in]{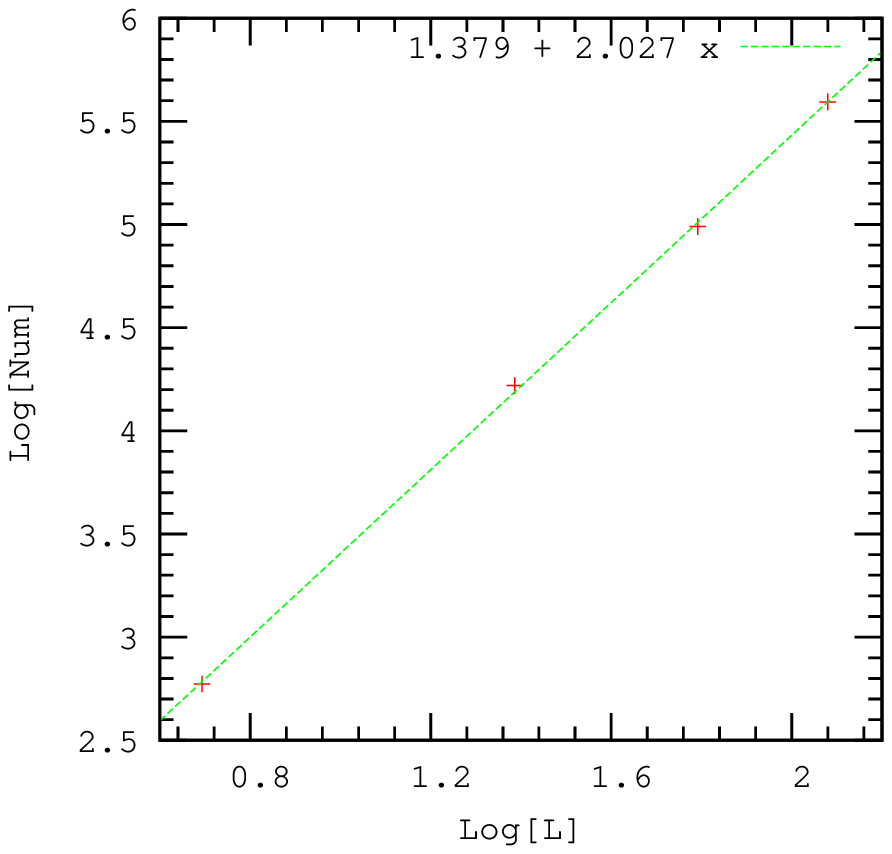}
    \includegraphics[width=3.5in]{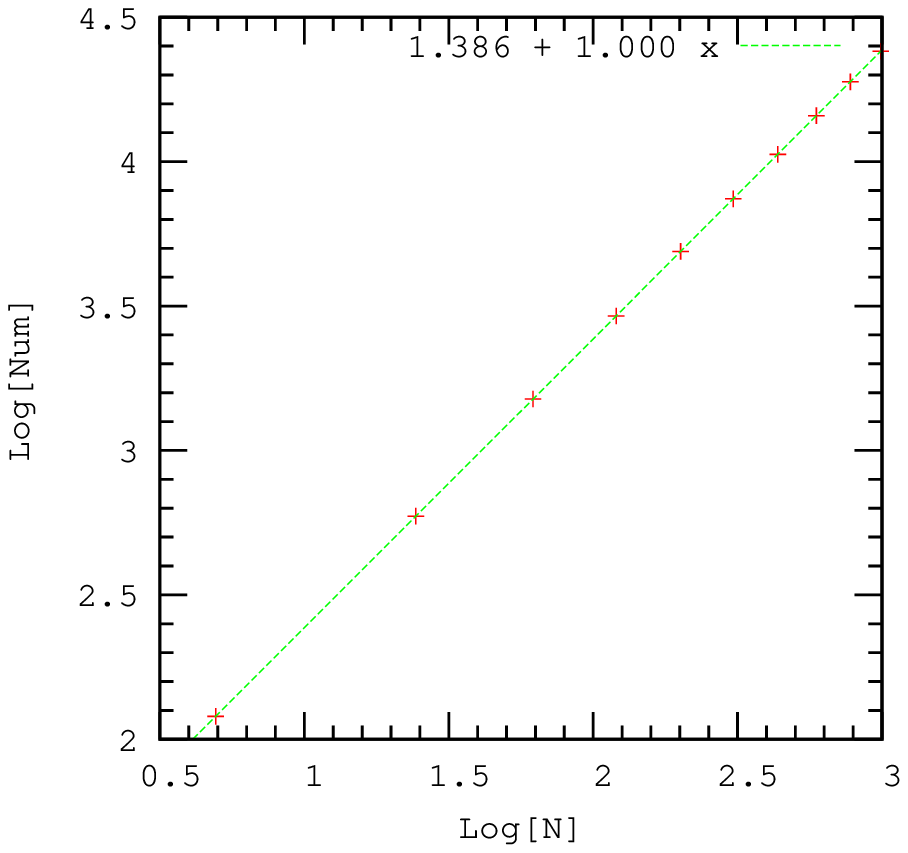}
		\caption{
				\label{fig:fitz}Number of zeros in a fixed region of the $b$-plane for $N=2$, $L$ variable (top) and $L=2$, $N$ variable (bottom)
}
	\end{figure}

The density of zeros increases with $N$ and $L$. We calculated the density of zeros in the $b$ plane (number of zeros in a given area of the $b$ plane taken as large as possible). The results are shown in Fig. \ref{fig:fitz}.  The fits of these log-log plots show that at fixed $L=2$, the density grows like $N^{1.000}$ and at fixed $N=2$, the density grows like $L^{2.027}$. This data is consistent with the idea that the density of zeros increases like the number of fields ($NV$).

\section{Density of State}
\label{sec:density}
Another way to obtain information about the location of the Fisher's zeros is to calculate the density of states $n(E)$ in the 
complex energy plane. First, we consider the case where the energy $E$ is real. 
The density of states $n(E)$ is the inverse Laplace transform of the partition function:
\begin{equation}
\label{eq:density}
n(E)=\frac{N}{2\pi i}\int _{K-i\infty}^{K+i\infty}{db e^{b N E}Z(b)} \  .
\end{equation}
The contour of integration is a vertical line in the complex $b$ plane with a constant positive real part $K$ otherwise arbitrary. 
For $L$ even, the relation between $Z(b)$ and $Z(-b)$ given in Eq. (\ref{eq:dual}) implies that 
\begin{equation}
\label{eq:dualn}
n(4V-E)=n(E) \  .
\end{equation}
We can now use the form of the partition function given in Eq. (\ref{eq:bk}) to obtain an explicit form. 
The only poles of the integrand are at $b=0$. If $E+\frac{V}{2}M^2_j>0$, we can close the contour by adding a semi-circle at infinity to the left and calculate the residue of the pole of order $i$ at $b=0$. 
If $E+\frac{V}{2}M^2_j<0$ we can close the contour by adding a semi-circle at infinity to the right and the closed contour includes no poles. 
Since all the $M^2_j$ are real and negative, it is clear that $n(E)=0$ for $E<0$. For $L$ even, Eq. (\ref{eq:dualn}) implies that  that $n(E)=0$ for $E>4V$. 
The final result for $E$ real and $L$ and $N$ even is that the density of states is 
piecewise polynomial for $0\leq E \leq 4V $ and zero outside this interval. An explicit form will be given in Eq. (\ref{eq:dos}).

\begin{figure}[h]
 \begin{center}
	\includegraphics[width=3.5in]{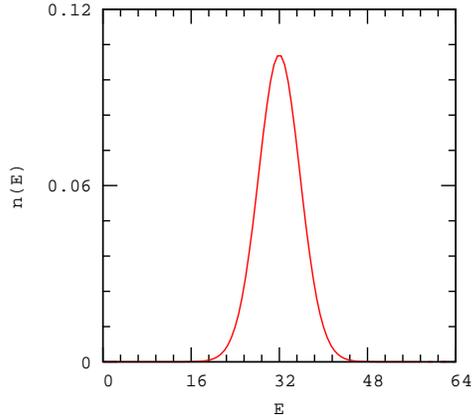}
\end{center}
		\caption{
				\label{fig:densitystate}Density of state function for $L=4$ and $N=2$
}
	\end{figure}

We now generalize this construction to the case where $E$ is complex. If we consider ${\rm e}^{bNE}$ on a circle at infinity in the $b$ plane, the expression blows up on one-half of the circle and decays on the other half. If we insist on being able to define the density of states by integrating Eq. (\ref{eq:bk}) term by term then the only way to extend the definition is to rotate the line integral in such a way that 
$bE$ is purely imaginary at both ends. The argument about the closing of the contour goes as before and we enclose the poles at 0 if $Re(E+\frac{V}{2}M^2_j)>0$. 
The final result is: 
\begin{equation}
\label{eq:dos}
n(E)=N \sum_{i,j}{a_{ij}\frac{(N E+\frac{N V}{2}M^2_j)^{i-1}}{(i-1)!} \theta(Re(E+\frac{V}{2}M^2_j)) } \  .
\end{equation}
Fig.\ref{fig:densitystate} shows the function for $L=4$ and $N=2$ following an explicit formula given in  Eq. (A.3) in the appendix. 

In summary, we have constructed a complex extension of the density of states that is stripwise polynomial in the complex $E$ plane. 
As the polynomials associated with $n(E)$ in two contiguous strips are different, it is unavoidable that some of their derivatives will be different at the boundary. 
However the discrepancies only appear at some order that increases with $VN/2$. This can be seen from the discussion in Section \ref{sec:model}, where we discuss the range of powers of $1/b$ appearing in the partition function. As we cross the boundary of a strip $Re(E)=V|M_j^2|/2$, 
we add  terms of the form $(E+\frac{V}{2}M^2_j)^{i-1}$ which vanish at the boundary if $i>1$. This term generates a discontinuity in the $i-1$-th derivative. The lowest value of $i$ occurring is $(V-n_j)N/2$, where $n_j$ is the number of times the pole $M_j^2$ appears. Consequently, the 
lowest derivative at which a discontinuity occurs is $(V-Max_j(n_j))N/2 -1 $. For instance,  for $L=4$, $Max_j(n_j)=6$, and the lowest order is $5N-1$. 
It can indeed be proven that for $L$ even, $Max_j(n_j)=2(L-1)$ and that it corresponds to $M^2_j=-4$ for which ``mirror" momenta can be paired 
in maximal number. 

Interestingly, we can use the large-$N$ limit to obtain finite volume thermodynamics from the density of states. 
At large $N$, by saddle point approximation of the Laplace transform of Eq. (\ref{eq:density}), we obtain that 
\begin{equation}
\label{eq:bofe}
b(E)\equiv f'=\frac{n'(E)}{N n(E)}\  , 
\end{equation}
which is similar to the standard thermodynamical relation $\beta=\partial S/\partial E$.

Eq. (\ref{eq:bofe}) provides a mapping of vertical strips of the $E$ plane into the $b$-plane. 
The solutions of $f''(E)=0$ give the singular points of this mapping. These singular points can be found 
strip by strip by finding the solutions of the polynomial equation: 
\begin{equation}
n(E)n''(E)=(n'(E))^2 \ ,
\end{equation}
that belongs to that vertical strip (assuming that $n(E)\neq 0$). 
We can then compare the $b$-images of these singular points using Eq. (\ref{eq:bofe}) strip by strip. 
Fig. \ref{fig:zero6} shows the images of these singular points for $L=6$, $N=2$ together with the images of the singular points of $b(M^2)$ and the zeros of partition function in $b$ plane. We see that the singular points of $b(E)$ seem to cluster is the same region as the singular points of $b(M^2)$ but with a slightly broader range. We conjecture 
that in the infinite volume limit, both types of singular points will accumulate on the outside boundary of the crossed shaped curve displayed in Fig. \ref{fig:l32} (image of the cut 
in the infinite volume limit). 

\begin{figure}
 
	\includegraphics[width=3.5in]{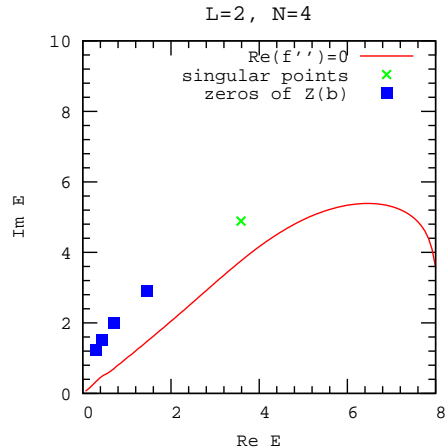}
		\caption{
				\label{fig:zsine} Singular points, zeros of partition function, and $Re(f'')=0$ in $E$ plane for $L=2$,$N=4$. The line (red online), corresponds to $Re(f'')=0$ with $Re(f'')>0$ above it. 
}
\end{figure}

In the saddle point approximation, it can be argued \cite{Bazavov:2009wz}, that Fisher's zeros can only appear as images of regions in the $E$ plane where $Re(f'')>0$. In Fig. \ref{fig:zsine}, we see that the inverse image of the Fisher's zeros appears in the region $Re(f'')>0$ even at small values of $N$ and $L$. 

\section{Complex RG flows for $O(N)$}
\label{sec:flow}

In this section, we construct complex RG flows using two different methods. The first one is based on a rescaling of the UV cutoff in $b(M^2)$ given by the saddle point relation Eq. (\ref{eq:gap}), the second is a complex extension of the two-lattice matching proposed in Ref. \cite{PhysRevB.27.1736,Hasenfratz:1984hx}

\subsection{Rescaling of $M^2$}

Eq. (\ref{eq:gap}) can be interpreted as a relation between the bare coupling and the UV cutoff keeping the renormalized mass, or mass gap, fixed. 
As the UV cutoff is lowered, the coupling increases and ultimately, $b$ flows to zero. 
More specifically, $M^2=m_R^2/\Lambda^2$ and we follow the change in $b$ under the change $\Lambda\rightarrow \Lambda/s$. In the literature 
the rescaling factor $s$ is often denoted $b$, but we are already using this symbol for the inverse 't Hooft coupling. Under this change, $M^2\rightarrow s^2 M^2$ and we can follow the trajectory in the $b$ plane. In the following, we will iterate the transformation with $s=2$ in order to allow a comparison 
with the other method for which $s=2$ is the simplest possibility. The initial values of $M^2$ were taken on a small circle around the origin in the $M^2$ plane and then multiplied repeatedly by 4. We can visualize these trajectories as ``rays" coming out of the origin in the $M^2$ plane. 

\begin{figure}[h]
 \begin{center}
   \includegraphics[width=2.8in]{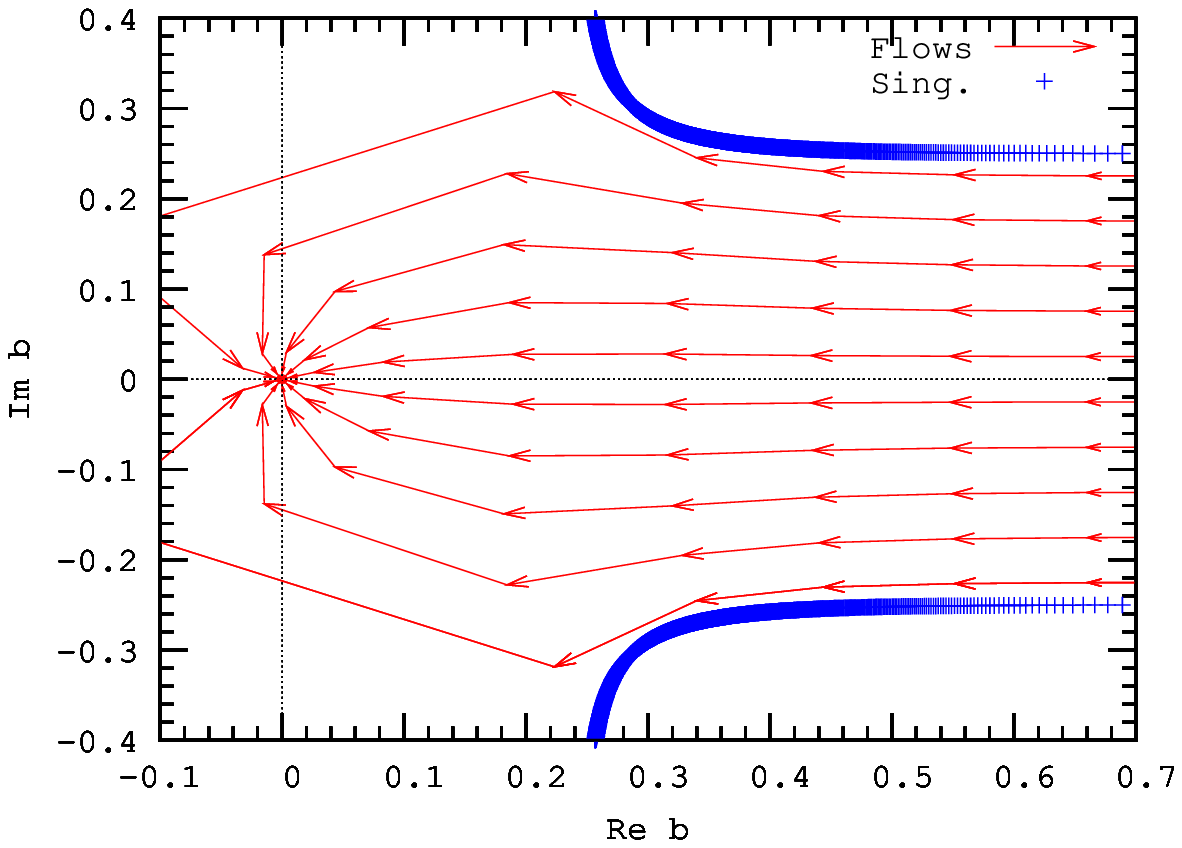}
	\includegraphics[width=2.8in]{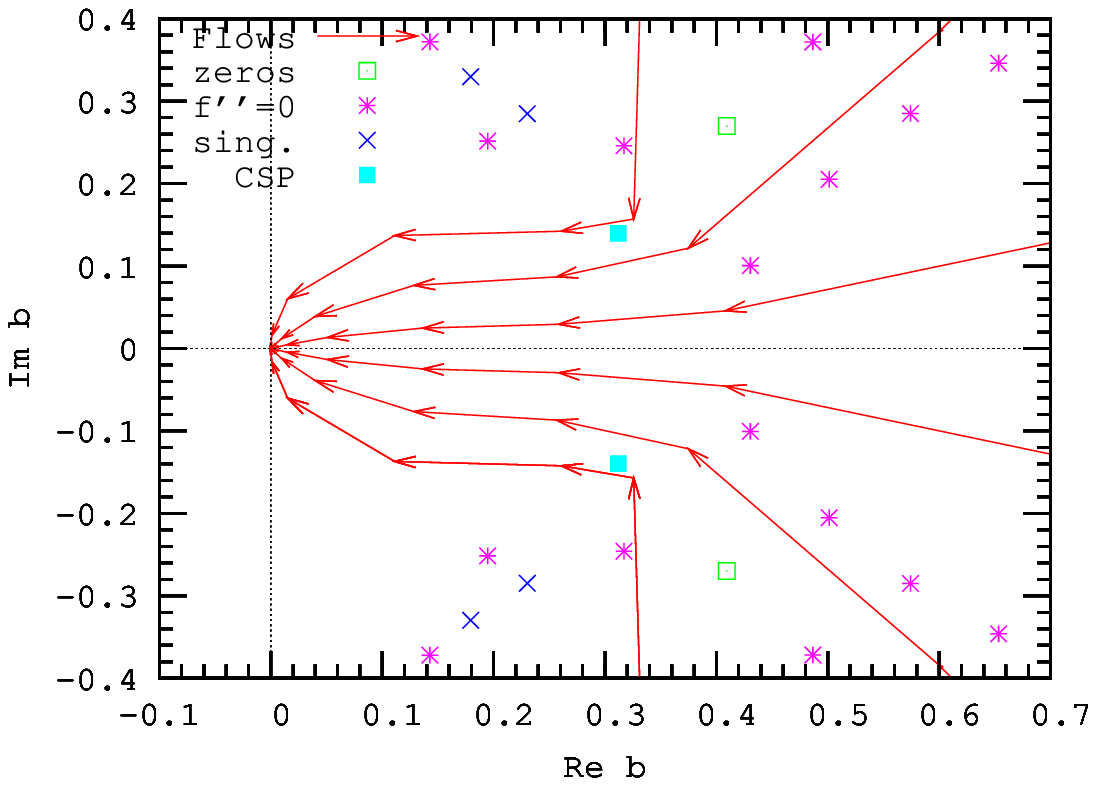}
	\includegraphics[width=1.9in,angle=270]{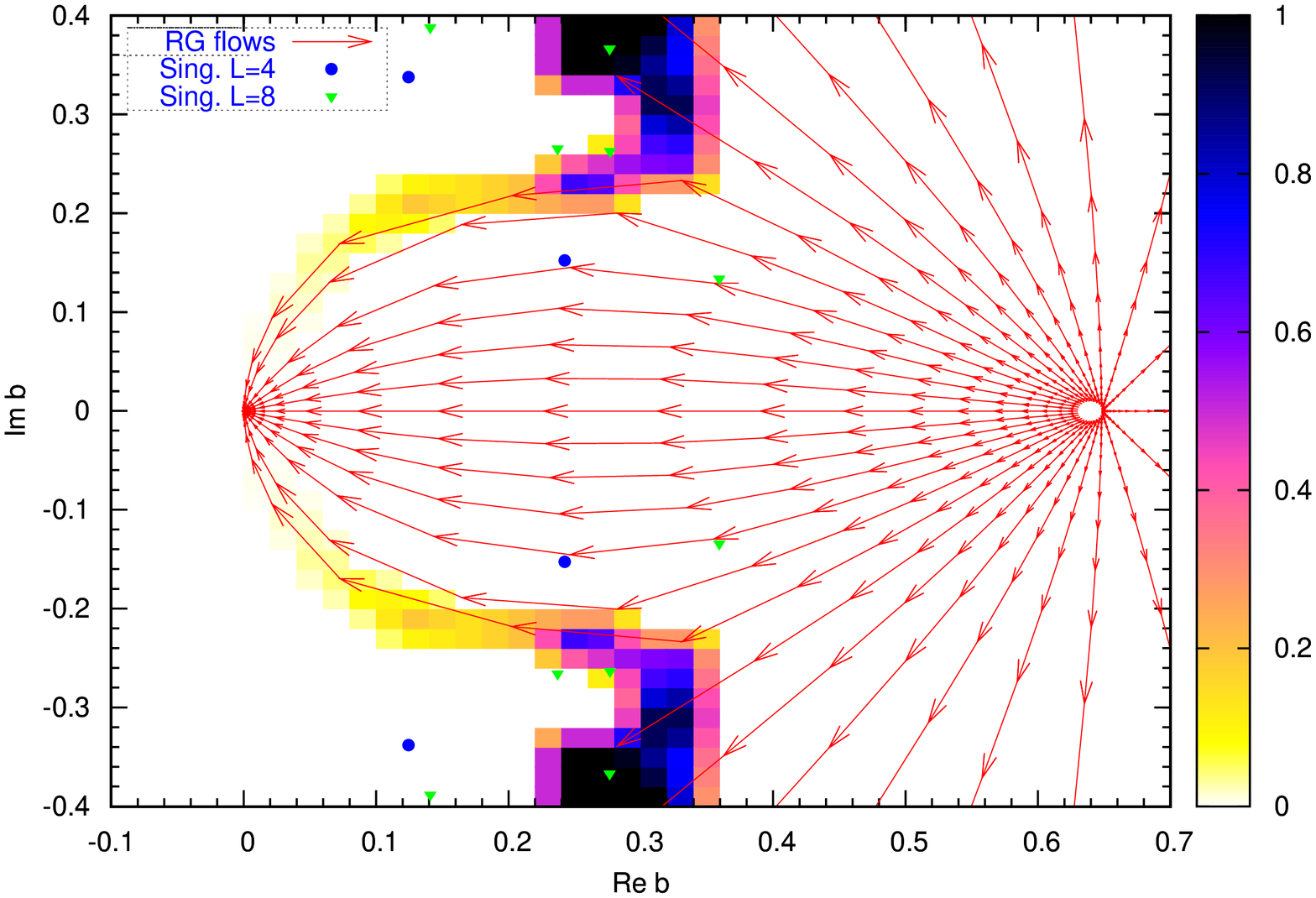}
\end{center}
		\caption{
				\label{fig:rescaling}a) (up) RG flows by rescaling and image of the cut at  infinite volume;   b) (bottom) flows by rescaling, singular points, CSPs,  
				 zeros of partition functions ($N=2$), and $f''=0$ ($N=2$) in $b$ plane for $6 \times 6$ lattice system.  c) RG flows for the 2-lattice matching between $8\times 8$ and $4\times 4$ lattices. Circles  and  triangles are the singular points for $L=4$ and $L=8$.
}
	\end{figure}

At infinite volume, as long as the trajectories in the $M^2$ do not cross the cut, the corresponding trajectories in the $b$ plane will stay inside the 
cross shaped image of the cut shown in Fig. \ref{fig:l32}. Sample trajectories are shown on Fig. \ref{fig:rescaling} a) and illustrate this idea. 
The flatness of the flow at larger values of $Re b$ can be understood from the approximate logarithmic scaling of $b$ and the fact the rescaling 
factor is real and does not affect the phase of the rays. 

At finite volume, the cut acquires a thickness and a structure described in Sec. \ref{sec:gap}. If the ray crosses the oval shaped regions on two sides of the singular point (see Fig. \ref{fig:cut}), then the corresponding trajectory in the $b$ plane will wrap around the image of the singular point in the $b$ plane 
using another Riemann sheet. At very small $M^2$, we have $b(M^2)\simeq 1/(L^2M^2)$ (instead of  a logarithmic dependence) and we have inverted rays at infinity in the $b$ plane. 
By small, we mean that the other poles at $M^2_j$ can be neglected which occurs if $M^2<<1/L^2$, in other words, if the Compton wavelength is larger than the volume of the system. Sample trajectories are shown in Fig. \ref{fig:rescaling} b) for $L=6$. The two singular points which have the closest distance to the real axis are called the closest singular points (CSPs).  

\subsection{Two-lattice matching}

The 2-lattice matching \cite{PhysRevB.27.1736,Hasenfratz:1984hx} is another method that can be used to  obtain complex RG flows. We will be matching observables $R(b, L)$  from systems with different lattice spacing but equal physical sizes. Under a RG transformation, the lattice spacing increases ($a \rightarrow s a$) and  the number of sites decreases ($L\rightarrow L/s$), but the physical length $La$ stays constant. In the following, we compare a $2^n\times 2^n$ lattice with 't Hooft coupling $b$ with a $2^{n-1} \times 2^{n-1}$ lattice with 't Hooft coupling $b'$. The $2^n\times 2^n$ model is blocked $n-1$ times while the coarser lattice model is blocked $n-2$ times.  
For an arbitrary lattice with $L=2^q$, we define the ratio of correlations of block observables (sum of all the spins inside  a $L/2\times L/2$  block $B$ or its nearest neighbor block  $NB$). 
\begin{equation}
\label{eq:rat}
R(b,L)\equiv\frac{\left\langle (\sum_{x\in B}\vec{\phi}_x )(\sum_{y\in NB}\vec{\phi}_y)\right\rangle_{b,L}}{\left\langle (\sum_{x\in B}\vec{\phi}_x)(\sum_{y\in B}\vec{\phi}_y ))\right\rangle_{b,L}} \  .
\end{equation}
Note that $R(b,L)$ is independent of the rescaling of the fields that needs to be performed in order to get an explicit form for the RG transformation. Note also that in order to define the numerator and denominator separately, we need to divide by the partition function, but these normalization factors cancel in the ratio. In the large-$N$ limit, the correlations are the same as for a Gaussian model 
with a mass $M^2$ defined as a function of $b$ by Eq. (\ref{eq:gap}). By using the binary decomposition of the integers between 1 and $2^{n-1}$, 
in the Fourier decomposition of the two-point function, we obtain 
\begin{eqnarray}
\label{eq:cor}
\left\langle (\sum_{x\in B}\vec{\phi}_x )(\sum_{y\in NB}\vec{\phi}_y)\right\rangle_{b,2^n}&=&\sum_{{\bf k}} {\rm cos}(2^{n-1}k_1){\mathfrak H}({\bf k}){\mathfrak G}({\bf k})\ , \cr
\left\langle (\sum_{x\in B}\vec{\phi}_x )(\sum_{y\in B}\vec{\phi}_y)\right\rangle_{b,2^n}&=&\sum_{{\bf k}} {\mathfrak H}({\bf k}){\mathfrak G}({\bf k})\  ,
\end{eqnarray}
with 
\begin{equation}
{\mathfrak H}({\bf k})=\prod_{l=1}^{n-1}[(1+{\rm cos}(2^l k_1))(1+{\rm cos}(2^l k_2))\ ,
\end{equation}
and the lattice propagator
\begin{equation}
{\mathfrak G}({\bf k})=1/[2(2-{\rm cos}(k_1)-{\rm cos}(k_2))+M^2] \ .
\end{equation}
After performing the sums over the momenta, $R(b,2^n)$ reduces to a ratio of polynomials in $M^2$ understood as function of $b$. 
Examples are given in Eqs. (A.4-7) in the appendix. As explained in Sec. \ref{sec:gap},  there are $q+1$ values of $M^2$ corresponding to one values of $b$. In the following, we always select the value corresponding to the ``main sheet". This includes the circle at infinity where the conventional strong coupling behavior applies.  In other words, we exclude the $q$ other values of $M^2$ corresponding to the small oval shaped region surrounding the cut.  

A complex RG map  can be constructed as follows. Given an initial complex value of $b$, we determine $M^2(b)$ corresponding to the main sheet. Using Eqs. (\ref{eq:rat}-\ref{eq:cor}), this result in a unique numerical value for $R(b,2^n)$. We then match this number with 
$R(b',2^{n-1})$ expressed as a ratio of polynomials in ${M'} ^2 $.  This results in a certain number of solutions for ${M'} ^2$. We only keep the ones corresponding to the main sheet. Each of these selected solutions determines a unique value of $b'$. If more than one remain, we only keep the one 
closest to $b$.    In order to quantify the level of ambiguity associated with this choice, we define the ambiguity $A$ as 
\begin{equation}
\label{eq:unambiguous}
A = \frac{|b- b'|_{min}}{|b -b'|_{next-to-min}} \  ,
\end{equation}
where $|b- b'|_{min}$ is the distance between $b$ and the closest solutions and ${|b -b'|_{next-to-min}}$ the distance to the next 
closest solution. 
If there is only one solution $A=0$ (${|b -b'|_{next-to-min}}=\infty$). 
There is also the logical possibility that none of the ${M'}^2$ solutions correspond to the main sheet, but we never encountered this case in practical calculations. 
In Fig. \ref{fig:rescaling}c), light regions correspond to small values of A while dark-colored regions stand for values of $A$ close to 1 (maximal ambiguity). 
When $|b- b'|_{min} \simeq |b -b'|_{next-to-min}$ it is not clear that we can define a RG flow, and not surprisingly chaotic behavior is often observed in these circumstances. 
 Unambiguous flows tend to stay in light-colored regions while ambiguous flows go into the dark regions and jump to other fixed points.  By looking at all the singular points, we find that the closest singular points (CSP) are close to the edges of dark-colored regions, which gives us the idea that the unambiguous RG flows are bounded by the singular points and the features of all the RG flows are controlled by the CSP. 
 Fig. \ref{fig:rescaling} shows the RG flows starting from the first fixed point on the real positive axis($b_0=0.64$) with the singular points of $4\times 4$ and $8\times 8$ lattice systems. 
 
\begin{figure}[h]
 \begin{center}
	\includegraphics[width=3.8in]{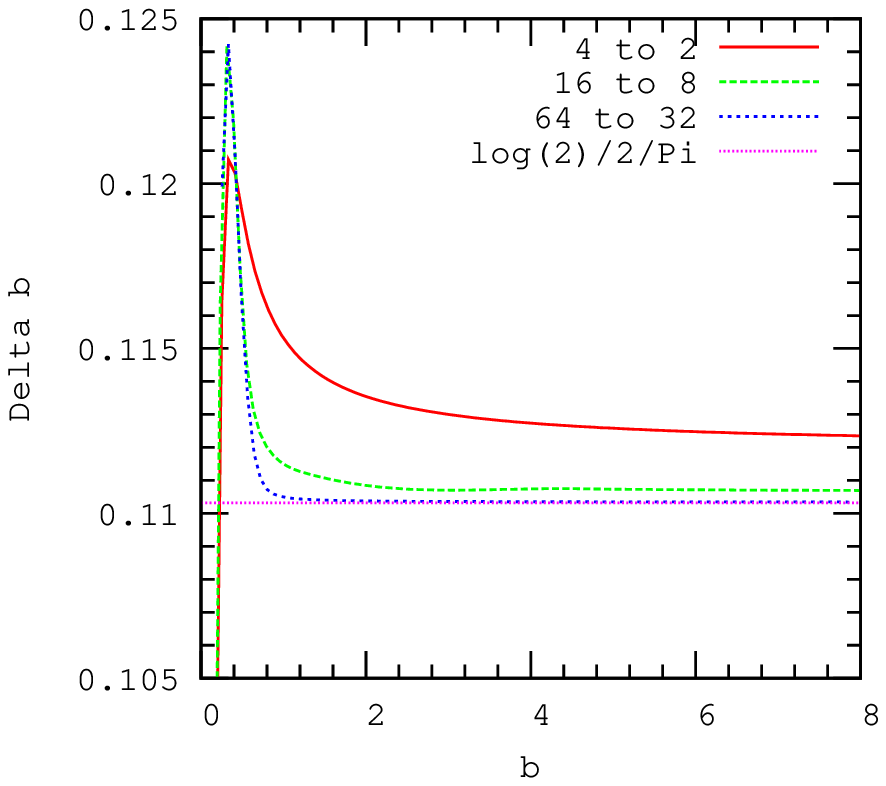}
   \includegraphics[width=3.8in]{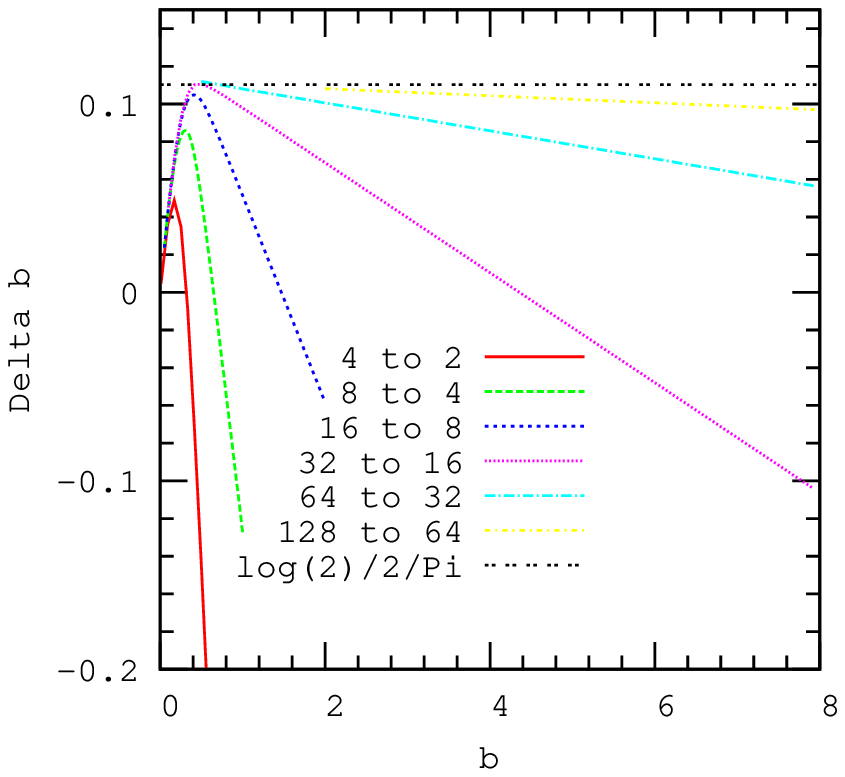}
\end{center}
		\caption{
				\label{fig:deltab}$\Delta b$ versus $b$ from rescaling (top) and 2-lattice matching (bottom).
}
	\end{figure}

In order to understand the fixed points on the real axis more systematically, we study the RG transformation in the small $M^2$, large $b$ limit. 
For very small $M^2$, the pole at zero dominates in Eq. (\ref{eq:gap})  and we have $b\simeq1/(M^2L^2)$ . The other poles have small contributions provided that  $M^2<<1/L^2$, in other words, when the Compton wavelength is larger than the linear size of the system. In the same limit, the matching condition  becomes 
\begin{equation}
{M'}^2\simeq 4(1-A/L^2) M^2\  , 
\end{equation}
with $A$ a constant that we will determine later. 
We now only consider the real solutions of $b$ and calculate the change of the coupling $\Delta b\equiv b-b'$. 
Numerical results are shown in Fig. \ref{fig:deltab} where a comparison with the rescaling method is made. 
In the infinite volume limit, the rescaling from $M^2$ to $4M^2$ leads to the relation:$\Delta b=\frac{1}{2\pi}\ln 2$ and for large $b$, there is no nontrivial fixed point. 
Putting everything together, we get the approximate finite volume formula:
\begin{equation} 
\label{eq:deltab}
\Delta b(b)\simeq-(A/L^2)b+\frac{\ln 2}{2\pi} \ .
\end{equation}
This implies the that we have an approximate fixed point 
\begin{equation}
b^\star_{app.}\equiv \frac{\ln 2}{A2\pi} L^2
\end{equation}
In Table.\ref{tab:compare}, we compare the numerical fixed points $b^\star_{num.}$ with the approximate  $b^\star_{app.}$ with $A=30.5$, and find the approximate model provides reasonable estimates for large volume. For comparison, $\Delta b$ can also be calculated with the rescaling method. In this case ${M'}^2=4M^2$ and if $M^2$ is finite and nonzero, the denominator of all the terms is $b'$ are strictly larger than for $b$, consequently $b'<b$ and there is no nontrivial fixed point. In summary, we see that discrepancies between the two RG methods occur in the limit where the Compton wavelength is larger than the size of the system. Otherwise, the RG flows going through the CSP are very similar for the two methods. 
\begin{table}
\begin{center}
\begin{tabular}{||c||c||c||}
\hline
$L$ & $b^\star_{num.}$ & $b^\star_{app.}$ \cr
\hline
4 & 0.320 & 0.058\cr
8 & 0.648 & 0.23\cr
16 & 1.47 & 0.93 \cr
32& 4.36 & 3.70 \cr
64 & 15.5 & 14.82 \cr
128 & 59.9 & 59.3\cr
256 & 237 & 237\cr
\hline
\end{tabular}
\end{center}
\caption{\label{tab:compare} Numerical solutions of fixed points and approximate solutions from $\Delta b = 0$ }
\end{table}

\subsection{Comparison of the two methods}

In this subsection we argue that in the infinite volume limit, the two RG flows discussed in the two previous subsections should coincide. By construction, the infinite volume of the RG flows illustrated in 
Fig. \ref{fig:rescaling} b) turn into those of Fig. \ref{fig:rescaling} a). It is nevertheless interesting to figure out in detail how it occurs. As we take initial conditions corresponding to 
$M^2=\epsilon {\rm e}^{i\theta}$, we see that as $\theta$ becomes slightly larger than $\pi/2$, the linear RG trajectories in the $M^2$ plane cross the rightest oval shaped region discussed in Sec. \ref{sec:gap} on the both sides of the upper singular  point. As a consequence, the corresponding  RG flows in the $b$-plane wrap around the CSP which is the image of the  first singular point, before going to 0. In the infinite volume limit, the CSP moves to infinity and we recover Fig. \ref{fig:rescaling} a). In some sense, the finite volume provides a 
regularization and gives a mathematical meaning to what happens for initial conditions corresponding to the cut.  

The two-lattice matching relies on the fact that if we apply the RG transformation enough times, the RG flows are projected on the unstable manifold which is one-dimensional 
for the models considered here. In this limit, the flows from the two lattices can be compared unambiguously. From the way we set up the calculation, it is clear that we reach an 
infinite number of RG transformation in the infinite volume limit 
(we block-spin ${\rm ln}_2(L)$ times).
 In this limit, the finite volume fixed point goes to infinity like $L^2$ and it is plausible 
that the unambiguous flows of Fig. \ref{fig:rescaling} c) become similar to those of  Fig. \ref{fig:rescaling} a). 

\section{Conclusions}

In summary,  we have extended two types of  RG transformations to complex coupling spaces for $O(N)$ models on $L\times L$ lattices in the large-$N$ limit. 
The three graphs of Fig. \ref{fig:rescaling} illustrate possible outcomes in models where calculations are more difficult such as asymptotically free lattice gauge theories.
Our general expectation is that at infinite volume, almost horizontal flows come from the large $\beta$ region where the logarithmic scaling characteristic of asymptotic freedom holds. 
The flows are then funneled through a continuous boundary whose shape is a lattice artifact. For instance, if we had used a continuum model with a sharp cutoff, the boundary would be flat instead of the shape seen in Fig. \ref{fig:rescaling} a). The existence of this boundary reflects the fact that we try to extend the mapping between $M^2$ and $b$, we encounter a cut 
in the $M^2$ plane and the boundary is the image in the $b$ plane of a close curve tightly enclosing the cut. 

At finite volume, the continuous boundary is replaced by a loosely defined region where complicated or ambiguous trajectories are observed (see also Fig. 1 in Ref. 
\cite{Denbleyker:2010sv}). Empirically, this region seems to coincide with the region where the images of the singular points of the mappings $b(M^2)$ or $b(E)$, and the ends of strings of Fisher's zeros appear. As the volume increases, the number of zeros in a fixed area of the $b$-plane increases like the volume and $N$ and we believe that the Fisher's zeros become dense outside of the boundary mentioned above. Fig. 3 in Ref. \cite{Denbleyker:2010sv} suggests that it will also be the case in lattice gauge theory. 
Our numerical study supports the idea that by monitoring the lowest zeros of asymptotically free theories when the volume increases, we can determine if the RG flows reach the region where a mass gap is present or if instead a nontrivial IR fixed point is encountered. We are planning to investigate the scaling of the Fisher's zeros in $SU(3)$ $N_f$ flavors and also 
to investigate the question of the corrections to asymptotic scaling \cite{Wolff:1990dm} in the complex plane. 

The complex flows described here are in some sense the simplest possible ones and we expect similar complex flows for  $SU(2)$ lattice gauge theory in 4 dimensions.  In contrast, for  $U(1)$ lattice gauge theory in 4 dimensions it appears that the Fisher's zeros pinch the real axis as $L^{-x}$ \cite{Bazavov:2010xh} with $2\leq x\leq 4$. A more precise estimate for $x$ will be discussed in a forthcoming preprint \cite{abprogress}. In this case, we expect complex RG flows similar to those of the 3 dimensional Ising or $O(N)$  models where two phases are present. 

Complex flows for models with a conventional second-order 
phase transition have been constructed for the hierarchical model \cite{Denbleyker:2010sv}. More recently, it has been found that the qualitative behavior of the complex flows can be modified by lowering the adjustable 
parameter (usually denoted $c$) below the critical value ($c$=1) where a second order phase transition is possible. As $c$ reaches 1, the nontrivial fixed point moves to infinity. As $c$ is further lowered, a pair of complex conjugated nontrivial 
complex fixed  points appear \cite{hmpreprint}. In general, it seems likely that by considering actions with tunable parameters, it is possible to create interesting patterns for  the complex RG flows. For hierarchical models with slightly modified 
interactions, it is possible to create additional fixed points and study their effects on the discrete $\Delta\beta$ function. In particular, we would like to understand if by constructing a continuous Calan-Symanzik $\beta$  reproducing approximately the discrete one, it is possible to relate the nontrivial complex fixed points to the complex zeros of the continuous $\beta$ function. This will also be discussed in Ref. \cite{hmpreprint}. 

For the models considered here, a richer complex flow behavior could be obtained by adding new terms in the energy function and considering higher dimensional complex flows. 
One possibility that comes to mind is to add a term inspired by the Witten-Wess-Zumino term in the continuum. Multidimensional RG flows can exhibit intricate global properties sorted in Ref. \cite{Polonyi:2001se}. 
Their complexification appear to be a completely open field of investigation.

\begin{acknowledgments}

Part of this work started  during the workshop 
``New applications of the renormalization group
method in nuclear, particle and condensed matter physics" held at the Institute for Nuclear Theory, University of Washington, Seattle (INT-10-45W). 
Additional work was done at the Aspen Center for Physics in May and June 2010 during the workshop ``Critical Behavior of Lattice models". 
We thank the participants of these two workshops 
for stimulating discussions. 
This 
research was supported in part  by the Department of Energy
under Contract No. FG02-91ER40664.

\end{acknowledgments}

\appendix*
\section{Numerical examples}
The numerical values of $2q$ for $L$ up to 42 are given in Table \ref{tab:n}.
\begin{table}[t]
\begin{center}
\begin{tabular}{||c|c||c|c||c|c||c|c||}
\hline
$L$& $2q$ & $L$ & $2q$ & $L$ & $2q$ & $L$ & $2q$ \cr
\hline
2 & 4 & 16 & 80 & 30 & 220 & 44 & 528\cr
4 & 8 & 18 & 100 & 32 & 288 & 46 & 576\cr
6 & 16 & 20 & 120 & 34 & 324 & 48 & 580\cr
8 & 24 & 22 & 144 & 36 & 328 & 50 & 676\cr
10 & 36 & 24 & 148 & 38 & 400 & 52 & 728\cr
12 & 40 & 26 & 196 & 40 & 440 & 54 & 784\cr
14 & 64 & 28 & 224 & 42 & 472 & 56 & 840\cr
\hline
\end{tabular}
\end{center}
\caption{\label{tab:n} $L$ and $2q$}
\end{table}
We also provide the explicit form for the partition function, $\gf (M^2)$ and the density of states for $D=2$, $L=4$, $N=2$. 

\vskip20pt

\begin{widetext}
\begin{eqnarray}
\nonumber
Z(b)&=&-\frac{1576575 e^{-128 b}}{4722366482869645213696
   b^{15}} [3456 b \left(2097152 b^4+174080
   b^2+2655\right) e^{64 b}\\ \nonumber
&-&5 e^{128 b}+5+40 e^{96 b} \left(294912 b^3-202752
   b^2+53472 b-5255\right)\\ \nonumber
&+&40 \left(294912 b^3+202752 b^2+53472b+5255\right) e^{32 b}]\\
&=&1-64 b+\frac{35328 b^2}{17}-\frac{2326528 b^3}{51}+\frac{737607680 b^4}{969}+O\left(b^5\right)
\end{eqnarray}
\begin{eqnarray}
\nonumber
\gf_{4\times 4}(M^2)=\frac{1}{16}\Big(1/M^2 &+&4/(2 + M^2) + 6/(4 + M^2) \\ + 4/(6 + M^2) &+& 1/(8 + M^2)\Big)\ .
\end{eqnarray}
\noindent
\begin{eqnarray}
\nonumber
n(E)&=&\frac{5}{7968993439842526298112}[-644087808(E-48)^{11}\theta(E-48)\\ \nonumber
&-&73801728(E-48)^{12}\theta(E-48)-2994432 (E-48)^{13}\theta(E-48)\\
\nonumber
&-&42040(E-48)^{14}\theta(E-48)-10882507603968(E-32)^9\theta(E-32)\\
\nonumber
&-&32848478208(E-32)^{11}\theta(E-32)-12845952(E-32)^{13}\theta(E-32)\\
\nonumber
&-&644087808(E-16)^{11}\theta(E-16)+73801728(E-16)^{12}\theta(E-16)\\
\nonumber
&-&2994432(E-16)^{13}\theta(E-16)+42040(E-16)^{14}\theta(E-16)\\
&+&E^{14}\theta(E)-(E-64)^{14}\theta(E-64)]\  .
\end{eqnarray}
We also give the explicit form of the first $R(b,2^l)$ with $M^2$ understood as a function of $b$ as explained in Sec. \ref{sec:flow}.
\begin{equation}
R(b,2) = \frac{8+2 M^2}{8+8 M^2+(M^2)^2}  \  ,
\end{equation}

\begin{equation}
R(b,4) = \frac{2+M^2}{2+4 M^2+(M^2)^2}   \  ,
\end{equation}
and
\begin{equation}
R(b,8) = \frac{32+82M^2+54(M^2)^2+13(M^2)^3+(M^2)^4}{32+222M^2+314(M^2)^2+153(M^2)^3+30(M^2)^4+2(M^2)^5}\ .
\end{equation}

\end{widetext}

%

\end{document}